\documentstyle[12pt]{article}
\newcommand{\beq}{\begin{equation}}
\newcommand{\eeq}{\end{equation}}

\def\half{{\textstyle{1\over2}}}
\def\quart{{\textstyle{1\over4}}}
\def\fhalf{{\textstyle{1\over4}}}
\def\kap{{\textstyle{1\over{\kappa^2}}}}
\def\gap{{\textstyle{1\over{g^2}}}}
\def\half{{\textstyle{1\over2}}}
\def\thre{{\textstyle{1\over3}}}

\def\thalf{{\textstyle{3\over2}}}
\def\quart{{\textstyle{1\over4}}}

\def\twff{{\textstyle{1\over{3\cdot 2^4}}}}
\def\eigt{{\textstyle{1\over{2^3}}}}
\def\ninsix{{\textstyle{{{1}\over{3\cdot 2^6}}}}}
\def\ninsixp{{\textstyle{{{1}\over{3\cdot 2^7}}}}}
\def\nineig{{\textstyle{{{1}\over{3\cdot 2^8}}}}}

\def\p1half{{\textstyle{{{p+1}\over{2}}}}}

\def\23phalf{{\textstyle{{{23-p}\over{2}}}}}

\begin{document}
\thispagestyle{empty}
\begin{titlepage}

\bigskip
\hskip 3.7in{\vbox{\baselineskip12pt
}}

\bigskip\bigskip\bigskip\bigskip
\centerline{\large\bf
A New Proposal for Matrix Theory}

\bigskip\bigskip
\bigskip\bigskip
\centerline{\bf Shyamoli Chaudhuri
\footnote{Current E-mail: shyamolic@yahoo.com}}

\medskip
\medskip
\medskip

\centerline{214 North Allegheny Street}
\centerline{Bellefonte, PA 16823, USA}

\bigskip

\date{\today}

\bigskip\bigskip
\begin{abstract}
\noindent We explain the motivation and main ideas underlying our
proposal for a Lagrangian for Matrix Theory based on sixteen
supercharges. Starting with the pedagogical example of a bosonic
matrix theory we describe the appearance of a continuum spacetime
geometry from a discrete, and noncommutative, spacetime with both
Lorentz and Yang-Mills invariances. We explain the appearance of
large $N$ ground states with Dbranes and elucidate the principle
of matrix Dbrane democracy at finite $N$. Based on the underlying
symmetry algebras that hold at both finite and infinite $N$, we
show why the supersymmetric matrix Lagrangian we propose does not
belong to the class of supermatrix models which includes the BFSS
and IKKT Matrix Models. We end with a preliminary discussion of a
path integral prescription for the Hartle-Hawking wavefunction of
the Universe derived from Matrix Theory.
\end{abstract}
\noindent

\end{titlepage}

\section{Introduction}

\vskip 0.1in
The pre-eminent task facing string theorists of our time is finding the answer
to the question: \lq\lq What is String Theory?" \cite{whatis}. We need an answer
that is plausible, consistent with all of the known facts about weak-strong-dual
effective field theory limits of nonperturbative String/M theory, and that is both
mathematically and aesthetically satisfactory
\cite{mthy,stfldth,bfss,dvv,ikkt,malda,kth}.
Much work has already been devoted to matrix model frameworks in this context
and for reviews we refer the reader to \cite{wati}. In recent work,
we have proposed a rather different direction of research, and the aim of this
paper is to explain the basic features of our proposal.

\vskip 0.1in Our matrix framework is motivated by a beautiful
property of the nine-dimensional supergravities
\cite{townhull,flux,fnew} that has received insufficient attention
in the literature, in our opinion. In nine spacetime dimensions,
but not in ten or eleven, it is possible to describe the full
spectrum of Dbrane potentials, including the ten-dimensional IIA
cosmological constant, in a manifestly covariant Lagrangian. By
field redefinitions, and by the action of weak-strong and
target-space dualities alone, it is therefore possible to connect
all of the known nine-dimensional ground states of M theory. This
extends, of course, to $32$ supercharges; we have restricted to
ground states with sixteen supercharges because of our belief that
chirality is a fundamental property of perturbative string theory,
worthy of emphasis in any fundamental formulation of the
nonperturbative theory. Thus, a somewhat modified view of the
well-known \lq\lq Star" diagram \cite{polbook} that inspired our
matrix framework is as follows. We represent the moduli space of
theories with $N$$=$$16$ supercharges by a star with six vertices.
Place a theory and its $T_9$-dual at opposite vertices. Then,
going clockwise around the star, we have the type I string with
$32$ D9branes, the massive IIA string with $32$ D8branes, the
heterotic string with gauge group $E_8$$\times$$E_8$, the type
I$^{\prime}$ string--- the same as $M$-theory compactified on
$S^1$$\times$$S^1/{\rm Z}_2$, the massive IIB string with $32$
D9branes, and the heterotic string with gauge group $Spin(32)/{\rm
Z}_2$. The weak-strong coupling duals of these theories lie on the
same diagram. The fact that the nonperturbative theory has a
hidden eleven-dimensional nature is evident, but all of the
vertices of the star are nine-dimensional. The matrix action we
will propose can be identified in the large $N$ limit with any of
the six vertices of this star, upto appropriate field
redefinitions.

\vskip 0.1in Our proposed Lagrangian for Matrix Theory assumes a
theory based on sixteen supercharges. Spacetime is discrete, and
noncommutative, with a full $N^2$ degrees of freedom contained in
the $U(N)$ adjoint variable $e^{\mu}_a$ associated with each point
in space. An auxiliary tangent space introduced at each point in
spacetime is flat, and assumed to have Lorentzian signature, $(-,
+, \cdots , +)$. We identify the minimal $U(N)$ invariant matrix
Lagrangian for two independent adjoint variables: $e_a^{\mu}$ and
$A_{\mu}$, consistent with Lorentz and Yang-Mills invariance at
both finite, and infinite, $N$. Since the target symmetries acts
noncommutatively on the space of $U(N)$ matrices, provision of a
prescription for matrix ordering is crucial in this framework. We
should clarify that the supersymmetry partners of the adjoint
variables live in the fundamental $N$-representations of the
$SU(N)$ subgroup. Thus, unlike both the BFSS and IKKT matrix
models, and some proposed extensions, our matrix Lagrangian is not
a super-matrix model based on the supersymmetrization of $U(N)$.

\vskip 0.1in
In the large $N$ limit, all matrix variables assume diagonal form and play the
role of continuum fields. A continuum spacetime emerges and all of the basic
elements of a spacetime geometry, including covariant derivatives and the
geodesic equation, assume the form taken in Riemannian geometry. From a
consideration of the spacetime $\times$ internal symmetries alone, it is evident then
that the large $N$ limit of our matrix Lagrangian with sixteen supercharges coincides
with the spacetime Lagrangian of a string theory
in the zero slope limit: we hold $g_{\rm o}$, $g_c$, and $N \alpha^{\prime 1/2}$ fixed in
the matrix action, taking
$N$ to $\infty$. We should point out that the large $N$ limit is not unique.
The reason is that the $1/N$ corrections in the matrix Lagrangian will
provide $O(\alpha^{\prime})$
corrections to the spacetime Lagrangian. These need not, a priori, agree with the known
$\alpha^{\prime}$ expansion of the string Lagrangian. However, when we demand that expansion
about the large $N$ limit result {\em in a perturbatively finite and anomaly free spacetime
Lagrangian}, agreement with the string theory Lagrangian follows because those properties
are unique.
Finally, for more general quantum backgrounds of String Theory, with Mbranes,
NSbranes and extensions, the $1/N$ terms from Matrix Theory make predictions for the
unknown $\alpha^{\prime}$ corrections to these backgrounds. Neither is the large N
limit unique in these cases; holding one or more mass scale fixed, in addition to
$N\alpha^{\prime 1/2}$, in the large $N$ limit can give inequivalent effective field theory
limits. This is familiar from the extensive experience with AdS/CFT large $N$ duals \cite{malda}
and with the noncommutative large $N$ N=4 SYM field theory limits of type II string theory
\cite{seiwit,ncom}.

\vskip 0.1in
Our assignment of independent matrix variables to gauge and gravitational degrees
of freedom has its origin in the properties of perturbative string theory. At short
distances, the noncommutative nature of spacetime manifests itself in the overlap
rules for closely spaced Dbranes. At large distances, noncommutativity manifests
itself as the existence of an antisymmetric two-form tensor potential, the natural
potential of choice in a theory of one-dimensional objects. We emphasize that this has
little to do with whether or not there is an antisymmetric tensor background in {\em
our} four-dimensional world, just one of a myriad of ground states of this theory.
Indeed, phenomenological constraints suggest otherwise. But there is no question that
the symmetric metric two-form and antisymmetric tensor twoform enter String Theory at
a more fundamental level on equal footing. We will therefore preserve this equivalence
in our matrix framework, obtaining both from a first order formalism for matrix gravity
based on vierbeins.

\vskip 0.1in The duality between short and long distance manifestations of noncommutativity
also mirrors the well-known open-closed worldsheet duality of perturbative
string theory. We will pause here to clarify an unfortunate misconception that has
crept into the recent string literature, with the oft-made remark that \lq\lq gauge
theory contains gravity". This is incorrect: open strings do produce closed
strings at the loop level, but the renormalization of the
open and closed string couplings are known to have independent
origin. In terms of the worldsheet \cite{poltorus,ncom}, this is seen in the fact
that both the coincidence limit of massless vertex operators, {\em and the limit of
vanishing loop lengths}, contribute to the renormalization of the open string coupling,
but not the closed string coupling. Furthermore, in any open and closed string theory
there is always a subsector of pure closed string diagrams. Thus, in the absence of
supersymmetric nonrenormalization theorems, the tree-level
relation, $g_c$$=$$g_o^2$, receives non-trivial correction order-by-order in string
perturbation theory. In short, the short-distance degrees of freedom accounting for the
renormalization of open (gauge) and closed (gravitational) couplings have independent origin.
We have emphasized this point because, unlike both the BFSS and the IIB Matrix Models,
where gravity is purely a \lq\lq derived" effective interaction while gauge theory is
fundamental, the open and closed sectors of perturbative string theory contribute on
independent footing to the long distance effective interactions. We have taken care to
preserve this property in our matrix framework.

\vskip 0.25in As a prologue to discussion of our proposal for
Matrix Theory, we begin with the simpler and more pedagogical
example of a {\em Bosonic Matrix Theory}.\footnote{We should
clarify at the outset that this bosonic matrix model has no
relation to the conjecture put forth in \cite{bosm}.} The
gravitational and gauge degrees of freedom in our theory belong to
two independent $U(N)$ adjoint multiplets: $e^{\mu}_a$ and
$A^{\mu}$. Here, $\mu$ labels the directions in spacetime, and $a$
labels the coordinates in an auxiliary flat tangent space
introduced at every point in spacetime. This procedure gives a
natural prescription for matrix ordering; an additional
consequence is that the full nonlinear part of the gravitational
interaction is already present in the classical action. Finally,
diffeomorphism invariance is manifest. We describe the emergence
of a continuum spacetime and the basic elements of a spacetime
geometry, including covariant differentiation and the geodesic
equation, from this framework. Important steps are the definition
of a volume element, and the definition of partial derivative and
integration of matrix variables at finite $N$. Given these steps,
we can obtain expressions for the Riemann curvature, Yang-Mills
tensor, and the full tower of higher rank antisymmetric $p$-form
field strength tensors. Requiring, in addition, invariance under
all of the higher rank gauge invariances of the bosonic matrix
theory provides a natural prescription for the quantum matrix
action.

\vskip 0.1in
The supersymmetric matrix theory is a nontrivial extension of the bosonic theory
which is not to be confused with the supersymmetrization of a $U(N)$ matrix model,
also known as a super-matrix model. In our framework, matrix variables in the same
supermultiplet belong to distinct $U(N)$ representations. Specifically, the gravitational
and gauge degrees
of freedom belong to distinct adjoints, $e_{\mu}^a$ and $A^{\mu}$, while
their superpartners, $\chi^{\alpha}_{\mu}$ and $\psi^{\alpha}$, where $\alpha$ is
a spinor index taking values $1$, $\cdots$, $16$, belong in distinct fundamental
representations of $U(N)$. We will find that this assignment naturally enables
chirality in the matrix Lagrangian, which has sixteen supercharges at both finite,
and infinite, $N$.
The quantum matrix action will be determined as before by requiring invariance
under the full tower of higher rank gauge symmetries.
We will find that matrix Dbrane states demonstrate a remarkably simple
and elegant phenomenon
we refer to as {\em Dbrane democracy:} closure of the
finite $N$ matrix Lorentz algebra in any matrix theory ground state with Dpbrane
charge in the presence of Yang-Mills fields {\em requires} that the ground state
is simultaneously charged under
the full tower of antisymmetric matrix potentials with $p$$\le$$26$.\footnote{This
is distinct, although not unrelated, to the use of the term {\em $p$-brane
democracy} in \cite{town}.} We should note that Dbrane democracy has a beautiful
large $N$ remnant in the continuum theory, in the
form of mixed Chern-Simons couplings in the
Lagrangian when the one-form gauge symmetry is nonabelian.

\vskip 0.1in
We close with a matrix path integral representation for the Hartle-Hawking
wavefunction, pointing the way to a derivation of the Wheeler De Witt equation
for Matrix Theory. We note that a full treatment will require a clarified
understanding of the role of compact De Sitter-like 9-geometries in String/M
theory, a subject of active ongoing research \cite{desit}. We conclude with a
discussion of possible future directions of research coming out of our work.

\section{Bosonic Matrix Theory}

\vskip 0.1in
The fundamental variables in the bosonic matrix Lagrangian are objects living
in the $N^2$-dimensional adjoint representation of the unitary group $U(N)$.
Notice that although the individual components of a bosonic matrix take
value in the field of ordinary real
(complex) numbers, the matrix itself is a noncommuting object obeying the
rules of $U(N)$ matrix multiplication. Thus, the ordering of matrices
within a composite product of $U(N)$ matrices is of crucial importance.
This is especially important since we will need to project onto
particular tensor products of adjoint representations in order to give matrix
expressions for the {\em physical} variables, such as the symmetric
two-tensor, $g_{\mu\nu}$, antisymmetric two-form, $A_{\mu\nu}$,
and scalar, $\Phi$. These are the variables that will appear in the matrix
Lagrangian. Thus, an unambiguous prescription for matrix ordering is necessary
prior to any meaningful analysis of matrix Lagrangians.

\subsection{A Prescription for Matrix Ordering}

We will now make the case that an unambiguous prescription for the ordering of
individual matrices in a composite operator is given by requiring that each
transform simultaneously in an irreducible representation (irrep) of the
unitary group, $SU(N)$, and in the $SL(n,C)$ subgroup of the inhomogenous
finite $N$ Lorentz group in $d$$=$$2n$ dimensions. The construction of an invariant
matrix Lagrangian built out of composite Lorentz scalars then proceeds by the
Noether procedure, familiar from analogous manipulations in classical field theory.

\vskip 0.1in
We work in the first order formalism for Einstein gravity. The basic objects in
our matrix Lagrangian are the vierbein, $ e_{\mu}^a $, a square array of size $4n^2$,
each element of which is a $U(N)$ adjoint, and which is subject to $2n$ constraints,
$e^{\mu}_a e_{\mu b}$$=$$\eta_{ab}$. Next, we have the nonabelian vector potential,
$A_{\mu}$, a one-dimensional array of size $2n$, each element of which is a $U(N)$
adjoint. Notice that the origin of gravitational degrees of freedom, and of the
spacetime continuum, is distinct from the origin of the Yang-Mills sector in this
framework. The independent assignment of gauge and gravitational sectors in our
construction is directly motivated by the analogous property of perturbative open
and closed string theories, as explained in the Introduction.

\vskip 0.1in
The dimensionality, $d$$=$$2n$, of the auxiliary flat tangent space may be
left undetermined in the classical theory at first, allowing for the possibility
of bosonic matrix theories with an arbitrary number of noncompact dimensions in
the large $N$ continuum limit. We will assume, however, the Minkowskian signature
$(-,+,\cdots,+)$
for the tangent space, which is coordinatized by $d$ real-valued parameters, $\xi^a$,
and has box-regulated volume $V_{d}$. Associated with each point in tangent space
is a whole $d(d$$-$$1)$$N^2$ unrestricted variables contained in the vierbein,
encapsulating information about the background spacetime geometry of some large
$N$ ground state of the matrix theory.
In what follows, we will work in the first order formalism for
gravity. The symmetric twoform metric tensor is the composite,
$e^{\mu}_a e^{\nu}_b \eta^{ab}$, the antisymmetric
twoform potential is the composite $e^{\mu}_a e^{\nu}_b \epsilon^{ab}$.
Finally, the dilaton is the scalar, $e^{\mu}_a e^{a}_{\mu}$.
Given their $SL(n,C)$ assignments, the kinetic terms for the $U(N)$
matrix variables described above take the manifestly invariant form:
\begin{equation}
{\cal L} =
- \fhalf \gap e^{-\Phi} {F}^{\mu\nu} F_{\mu\nu} -
   \half \kap e^{- 2 \Phi} ( {\cal R}
              - 4 \partial^{\mu} \Phi  \partial_{\mu} \Phi  ) -
  \thalf \kap e^{-2\Phi} {H}^{\mu\nu\lambda} {H}_{\mu\nu\lambda}  \quad .
\label{eq:bmatc}
\end{equation}
Individual terms in the matrix Lagrangian are both $U(N)$ and Lorentz
scalars. The Lagrangian has been written in terms of the composite $U(N)$
variables with direct correspondence to the fields appearing in the low
energy
spacetime Lagrangian of string theory: the scalar dilaton, $\Phi$,
symmetric two-form or metric,
$g_{\mu\nu}$, and antisymmetric two-form, $A_{\mu\nu}$. At this juncture,
it would be helpful
to clarify how the spacetime continuum emerges from this framework in the
large $N$ limit. We must also give concrete meaning to the various matrix-valued
symbols in the Lagrangian. A crucial step will be the definition of matrix
partial differentiation and matrix integration. We will also clarify the
origin of spacetime symmetries such as Lorentz invariance and Yang-Mills
gauge invariance.

\subsection{Emergence of the Spacetime Continuum}

\vskip 0.2in
Let us put some intuition into the algebraic notions described
above by understanding how the spacetime continuum emerges in this
framework.
The basic idea is to give a
suitable definition of length, area, and volume valid in the non-continuum
finite $N$ case, clarifying simultaneously the notion of matrix partial
differentiation and matrix integration. The nature of the spacetime symmetries
such as Lorentx and Yang-Mills invariance outside of the large $N$ limit
serves as our guiding principle in arriving
at these definitions.

\vskip 0.2in
We introduce a continuum flat tangent space coordinatized by the
variables $\xi^a $, $a$$=$$0$, $\cdots$, $d$$-$$1$, at every point in
space. Spacetime itself is discretized, and there are $N^2$ degrees of
freedom associated with each coordinate rather than the expected $N$.
Thus, points in spacetime are in one-to-one correspondence with
matrices $e_{\mu a}$, where $X_{\mu}$$\equiv$$e_{\mu a} (X) d \xi^a$,
$\mu$$=$$0$, $\cdots$, $d$$-$$1$.
We will now give Lorentz invariant definitions for infinitesimal length
and area elements as follows. We define the
length of the $d$-dimensional position vector, $ X^{\mu}$, where $X$
is an $N$$\times$$N$ $U(N)$ matrix, as follows:
\begin{equation}
|X|^2 ~=~ {\rm Tr}_{U(N)} ~ e^{\mu}_{ a} e_{\mu b} d\xi^{a} d \xi^{b}  \quad ,
\label{eq:length}
\end{equation}
where the trace is over $U(N)$ indices. In performing concrete calculations, it
will be helpful to work in a proper time gauge in which
$X^0$ is taken to be diagonal, and the elements along the diagonal increase
smoothly, and monotonically, denoting time. In this gauge, we will identify
$X^0$$=$$\xi^0$.

\vskip 0.1in
In the large $N$ limit,
all of the spatial
$X^{\mu}$ will also take diagonal form, and the elements along the diagonal
of each $X$ will increase smoothly and monotonically denoting the coordinates
of space. Notice that the result of the trace is an ordinary real number
denoting the position of some event with respect to an arbitrarily
chosen origin. Translation of the origin corresponds to a $U(N)$
transformation. The interval
between two neighbouring events in spacetime,
$ds^2$, where $X^{\mu}$, $X^{\prime \mu }$$=$$X^{\mu}$$+$$(\Delta X)^{\mu}$, denote events
separated by the increment $(\Delta X)^{\mu}$,
is given by:
\begin{equation}
ds^2 ~=~ {\rm Tr}_{U(N)} ~ \Delta e^{\mu}_{ a} \Delta e_{\mu b} d\xi^{a} d \xi^{b}  \quad ,
\label{eq:lengthd}
\end{equation}
an invariant length for a given class of inertial observors.
The total length along some given curve, $\cal C$, in spacetime,
parameterized by a proper time, $\lambda$,
with respect to a
chosen inertial observor, is given by
the integral:
\begin{equation}
 l ~=~ \int_{\cal C} | {\rm Tr}_{U(N)} ~
    {{ \Delta e^{\mu}_{ a}}\over{d \lambda}}{{ \Delta e_{\mu b} }\over{d \lambda}}
   d\xi^{a} d \xi^{b} |^{1/2} d \lambda \quad .
\label{eq:lengthdl}
\end{equation}
The result for the length is, of course, identical for a class
of inertial observors and independent of the choice of
parameterization for proper time, or of affine parameter.

\vskip 0.1in
Similar definitions can be given for the
$p$-th volume form, $p$$=$$2$, $\cdots$, $d$. Begin with a local Lorentz
frame where the $d$-th volume element is simply given by:
\begin{equation}
dV ~=~ {\rm Tr}_{U(N)} ~ \Delta e^{0}_{ a_0} \cdots
  \Delta e^{d-1)}_{ a_{d-1}} d\xi^{a_0} \cdots d \xi^{a_{d-1}}
\quad .
\label{eq:voll}
\end{equation}
The result for the volume element in an arbitrary coordinate system follows:
\begin{equation}
dV ~=~ \left [ {\rm det} (-g)]^{1/2} \right ]
   {\rm Tr}_{U(N)} ~ \Delta e^{0 \prime }_{ a_0} \cdots
       \Delta e^{ (d-1) \prime}_{a_{d-1} } d\xi^{a_0} \cdots d \xi^{a_{d-1}}
\quad ,
\label{eq:volc}
\end{equation}
where $g$ is the metric tensor:
\begin{equation}
\left [  {\rm det} (-g)]^{1/2} \right ]
  =
  \left [ {\rm det}_{U(N)}~ (- e_{\mu a } e_{\nu b} \eta^{ab} )]^{1/2} \right ]
\quad .
\label{eq:metd}
\end{equation}
The covariant derivative is defined as follows. First, we write down
an expression for partial differentiation at a given point labelled
by the $U(N)$ matrix $X$ by referring to the differentials in the
local tangent space:
\begin{equation}
{{\partial}\over{\partial X^{\mu}}} \equiv
{{\partial}\over{\partial \xi^{c}}}
 | {{\Delta X^{\mu}}\over{\partial \xi^{c}}} |^{-1}
\quad .
\label{eq:partial}
\end{equation}
The inverse on the R.H.S. of this equation denotes taking the
$U(N)$ inverse of the infinitesimal matrix $(\Delta X)$. Matrix integration
will correspondingly be defined as multiplication by $\Delta X^{\mu}$,
such that:
\begin{equation}
\int_C (\Delta X^{\mu})^{-1} \Delta X^{\mu}
= 1 \quad ,
\label{eq:integ}
\end{equation}
and where the integration is understood to be path ordered.
The definition of the Christoffel
connection takes the form:
\begin{equation}
\Gamma^{\mu}_{\nu \lambda} \equiv
   \half g^{\mu\delta} \left (
 | {{\Delta X^{\lambda}}\over{\partial \xi^{c}}} |^{-1}
g_{\delta \nu,c}
+ | {{\Delta X^{\nu}}\over{\partial \xi^{c}}} |^{-1}
g_{\delta \lambda,c}
- | {{\Delta X^{\delta}}\over{\partial \xi^{c}}} |^{-1}
g_{\lambda \nu ,c}
\right )
\quad .
\label{eq:chris}
\end{equation}
The expressions for covariant differentiation follow. Specifically,
given Lorentz tensors $A^{\mu}$, $T^{\mu\nu}$, which are simultaneously
$U(N)$ matrices, we have:
\begin{equation}
A^{\mu}_{; \lambda}
\equiv
 | {{\Delta X^{\lambda}}\over{\partial \xi^{c}}} |^{-1}
   {{\Delta A^{\mu}}\over{\partial \xi^c}}
+ \Gamma^{\mu}_{\nu \lambda} A^{\nu}
,
\quad
T^{\mu\nu}_{; \lambda}
\equiv
 | {{\Delta X^{\lambda}}\over{\partial \xi^{c}}} |^{-1}
   {{\Delta T^{\mu\nu}}\over{\partial \xi^c}}
+ \Gamma^{\mu}_{\delta \lambda} T^{\delta\nu}
+ \Gamma^{\nu}_{\delta \lambda} T^{\mu\delta}
\quad .
\label{eq:covd}
\end{equation}

\vskip 0.1in
\noindent{\em Gauge Covariant Derivative:}
We now introduce a different $U(N)$ matrix variable, $A_{\mu}$, also carrying a
vector index. A diagonal configuration denotes a smoothly varying classical
field, the diagonal entries of which are ordinary continuous functions
of spacetime: $(A_{\mu})_{ab}$$=$$\delta_{ab} a_{\mu} (x_0, \cdots , x_{d-1})$.
In the proper time gauge, the distance along the diagonal will correspond
to the field's progression
in time. In particular, static or stationary configurations correspond to
a single non-vanishing diagonal element of $A$, a smooth function of the
spatial coordinates alone. In Minkowskian spacetime, the field
configuration as measured
with respect to an inertial observor's proper time will differ from that
measured by a different inertial observor: the difference is given
by the usual Lorentz transformation of the fields.
Finally, the vector potential also
carries charge under
the internal symmetry group, the Yang-Mills symmetry group, $G$. Thus,
the $U(N)$ adjoint is simultaneously a $d_G$-dimensional multiplet
under $G$. Thus, in flat Minkowskian spacetime, the gauge covariant derivative
takes the form:
\begin{equation}
D_{\lambda} A_{\mu}
\equiv
 | {{\Delta X^{\lambda}}\over{\partial \xi^{c}}} |^{-1}
   {{\Delta A_{\mu}}\over{\partial \xi^c}}
+ g [A_{\lambda} , A_{\mu}]
\quad .
\label{eq:gcovd}
\end{equation}
In a general curved
spacetime, we must use the Christoffel connection defined earlier to relate
the vector potential or field strength as measured by a non-inertial observor.

\vskip 0.1in
\noindent{\em Parallel Transport and Geodesics:}
Recall that a curve $\cal C$
in spacetime is a progression of $U(N)$ matrices, $\bf V$,
labelled by a parameter, the proper time $\lambda$. Thus, the tangent to the
curve is given by the progression of matrices, $\bf U$$=$$d {\bf V}/d\lambda$.
Here, bold-faced symbols denote $d$-vectors. Given
the basic elements
that describe the emerging geometry of the spacetime continuum, we can write
down the geodesic equation of motion for a test particle. Choose a local
inertial system at a given point ${\cal P}$ such that all components of
the Christoffel connection vanish at that point. It follows that
$\Delta V^{\mu}/d\lambda$$=$$0$$=U^{\nu}V^{\mu}_{;\nu}$ at $\cal P$.
This defines frame invariant parallel transport along the curve $\cal C$.
Finally, parallel transport of the tangent vector itself determines
the geodesics which satisfy
$\nabla_{\bf U} {\bf U}$$=$$0$, or:
\begin{equation}
U^{\mu} U^{\nu}_{;\mu}=U^{\mu}
 | {{\Delta X^{\mu}}\over{\partial \xi^{c}}} |^{-1}
   {{\Delta U_{\nu}}\over{\partial \xi^c}}
+ U^{\mu} \Gamma^{\nu}_{\mu\lambda}
 U^{\lambda} = 0
\quad .
\label{eq:geod}
\end{equation}

\vskip 0.1in
\noindent{\em Curvature:}
The result of parallel transport of a vector about an infinitesimal closed
loop in spacetime at a given point $\cal P$ gives the Riemann curvature
tensor, defined in the usual way, and appearing also in the commutator of
covariant derivatives:
\begin{equation}
R^{\mu}_{\nu\lambda\delta} =
\Gamma^{\mu}_{\nu\delta,\lambda}
-\Gamma^{\mu}_{\nu\lambda,\delta}
+\Gamma^{\mu}_{\sigma\lambda}
\Gamma^{\sigma}_{\nu\delta}
+\Gamma^{\mu}_{\sigma\delta}
\Gamma^{\sigma}_{\nu\lambda}
,
\quad [\nabla_{\mu} , \nabla_{\nu} ] V^{\lambda} = R^{\lambda}_{\delta \mu\nu} V^{\delta}
\quad .
\label{eq:curvd}
\end{equation}
Partial differentiation with respect to the $X$ is defined as above,
with reference to differentials in the local tangent space. The result is
the Bianchi identities.

\subsection{Point Sources and the Newtonian Potential}

\vskip 0.1in
In common with its predecessors \cite{bfss,ikkt}, the bosonic matrix theory
described above is a second quantized theory in target space, in the sense
that a classical matrix configuration may
describe one, or multiple, matrix objects. The simplest objects are point
sources; the
classical equation of motion for a free test particle is simply
the geodesic equation
given earlier. We will work in the proper time gauge setting $X^0$
equal to $\tau$$=$$\xi^0$. As in \cite{bfss,ikkt,wati}, in the nonrelativistic limit
we can consider a block diagonal configuration matrix
${\bf U}_i$$\equiv$${{d{\bf X}_i}\over{d\tau}}$, where $X_i(\tau)$
gives the location of the $i$th test particle:
\begin{eqnarray}
{\bf U} =
{\bf U}_1 & 0 & \cdots
\cr
0 & {\bf U}_2 & \cdots
\cr
\cdots & 0 & {\bf U}_n
\cr
\label{eq:points}
\end{eqnarray}
In this limit, the equation of motion is separable:
each test particle satisfies the geodesic equation,
$\nabla_{\bf U_i} {\bf U_i} (\tau) $$=$$0$, parameterized by
a common proper time, $\tau$, and taking the explicit form:
\begin{equation}
\left ( {{\Delta X^{\mu}}\over{d \tau}} \right )
 \left ( {{\partial}\over{\partial \xi^{c}}}
 | {{\Delta X^{\mu}}\over{\partial \xi^{c}}} |^{-1}
  \left [ {{\Delta X^{\nu}}\over{d \tau}} \right ] \right ) +
\left ({{\Delta X^{\mu}}\over{d \tau}} \right )
\Gamma^{\nu}_{\mu\lambda}
\left ( {{\Delta X^{\lambda}}\over{d \tau}} \right )
= 0 \quad .
\label{eq:pointgeo}
\end{equation}
Identifying $\tau/m_i$, where $m_i$ is the $i$th particle's
mass, as affine parameter, we can also express the geodesic
equations in terms of the momenta, ${\bf p}_i$$=$$m_i
d{\bf X}_i/d \tau$.

\vskip 0.1in
Next, we will extract the Newtonian potential sufficiently far from a
nonrelativistic point source in the linearized limit of the Einstein
equation $R_{\mu\nu}$$-$$\half R g_{\mu\nu}$$=$$8 \pi T_{\mu\nu}$. In
the limit of weak
disturbances, $g_{\mu\nu} $$=$$\eta_{\mu\nu}$$+$$h_{\mu\nu}$,
$|h_{\mu\nu}|$$<<$$1$, the curvature tensor takes the form:
\begin{equation}
R_{\mu\nu\lambda\sigma} =
h_{\mu\sigma,\nu\lambda}
+h_{\nu\lambda,\mu\sigma}
-h_{\mu\lambda,\nu\sigma}
-h_{\nu\sigma,\mu\lambda}
 \quad .
\label{eq:curvwf}
\end{equation}
Defining ${\bar{h}}^{\mu\nu}$$=$$h^{\mu\nu}$$-$$\half\eta^{\mu\nu} h$, in
the Lorentz gauge, ${\bar{h}}^{\mu\nu}_{\mu}$$=$$0$, we have the linearized
Einstein equations:
\begin{equation}
( -\partial_{\tau}^2 + \nabla^2 )
{\bar{h}}^{\mu\nu} = - 16 \pi T^{\mu\nu} \quad ,
\label{eq:einswf}
\end{equation}
The Newtonian limit applies when the disturbances are too small to attain
relativistic velocities. Thus, $|T^{00}|$$>>$$|T^{0i}|$$>>$$|T^{ij}|$,
which, from the linearized Einstein equation, implies that,
$|{\bar{h}}^{00}|$$>>$$|{\bar{h}}^{0i}|$$>>$$|{\bar{h}}^{ij}|$. The
dominant equation is that for the Newtonian potential,
$-4V$$\equiv$${\bar{h}}^{00}$, and upon setting
$T^{00}$$=$$\rho$$+$$O(\rho v^2)$, and dropping terms of
order $v^2 \nabla^2$, we have Newton's equation:
\begin{equation}
\nabla^2 V =  4 \pi \rho \quad .
\label{eq:newtwf}
\end{equation}
The far field potential for a stationary localized source
is given by the solution to
$\nabla^2 {\bar{h}}^{00}$$=$$0$, which is the Laplace equation. Its
solution in 26 spacetime dimensions takes the form,
$-C/r^{23}$$+$$O(1/r^{24})$. The coefficient can be determined on
dimensional grounds and by a matching calculation to the known
field theory limit. This is the static interaction, with $O(v^2)$
corrections in the bosonic matrix theory. In the 10d supersymmetric
case, as is well known, supersymmetry results in cancelation of both
the static and $O(v^2)$ corrections; the leading term in the
nonrelativistic potential is $O(v^4/r^7)$.

\vskip 0.1in
This completes our discussion of the linearized limit of the Einstein
equations. However, the full nonlinear part of the gravitational
interaction has already been included in the classical theory. Unlike the
case of M(atrix) Theory and the IIB Matrix Model, there is no need to
invoke quantum effects or the subtleties of the large $N$ limit
in order to account for the nonlinear part of the gravitational
interaction.

\subsection{Extended Objects from Matrices}

\vskip 0.1in
In addition to pointlike gravitational sources, the bosonic matrix
theory contains extended objects like Dbranes. Such extended matrix
theory objects couple to background antisymmetric tensor potentials
of higher rank, $C_{[p]}$, $p$$\le$$26$. In such a background, the
$SO(p$$-$$1$,$1)$ Lorentz subalgebra on the worldvolume of the
p-brane extends to an inhomogenous Lorentz algebra---
extended by the $\mu$$=$$0$, $\cdots$, $p$$-$$1$, hermitian
generators of spacetime
translations, giving the Poincare group in $p$ dimensions:
\begin{equation}
[ J_{\mu\rho} , J_{\nu \lambda} ] =
g_{\rho\nu} J_{\mu\lambda}
- g_{\mu\nu} J_{\rho\lambda}
- g_{\lambda\mu} J_{\nu\rho}
+ g_{\lambda\rho} J_{\nu\mu} ,
\quad [ P_{\mu} , J_{\nu \lambda} ] = g_{\mu\nu} P_{\lambda}
 - g_{\mu\lambda} P_{\nu}
\quad ,
\label{eq:poin}
\end{equation}
with $[ P_{\mu} , P_{\nu} ] $$=$$ 0$.
The $J_{\mu\nu}$ are the generators of spatial rotations, hermitian,
and antisymmetric in $\mu$, $\nu$.
All matrix generators have been denoted by $U(N)$ adjoints. In the classical theory,
the commutator denotes the Poisson bracket. It is replaced by the operator-valued
Heisenberg commutator in the quantum theory. The remnant $SO(26$$-$$p$,0)
Lorentz algebra coordinatizing directions orthogonal to the pbrane does not
extend to an inhomogenous algebra.

\vskip 0.1in
In the quantum theory, the commutators above are to be understood as
operator-valued symbols acting on a ground state with the required
properties. We begin with considering stationary, time independent, spacetime
geometries. Specifying such a ground state requires a stationary background
metric and stationary background gauge potentials. Begin with the spacetime
\lq\lq grid",
the family of $\{ e^{i}_a d \xi^a \} $ isomorphic to the spatial coordinates
$X^i$, for all time. In the large $N$ continuum limit, the $e_i^a$ are diagonal,
with entries along the diagonal displaying a smooth and monotonic increase in
the case of noncompact coordinates. For a compact coordinate, the diagonal
entries must display the required periodicity.
The volume element in the worldvolume of a p-brane takes the form:
\begin{equation}
dV = {\rm Tr} ~
\Delta e_{a_0}^0 \Delta e_{a_1}^1 \cdots \Delta e_{a_p}^p  d \xi^{a_0} \cdots d \xi^{a_p}
\quad .
\label{eq:volp}
\end{equation}
It is evident that the Poincare algebra given above acts as a set of operator
identities on the volume element in the worldvolume of the p-brane. The
same is true for any function of the $e^{\mu}_a$: the $\bf P$, $\bf J$, act
as derivatives on all such functions, where partial derivatives with respect to
$X^{\mu}$$\equiv$$e^{\mu}_a d \xi^a$ are as previously defined in section 2.2.
For example, we have the usual plane-wave basis for generic eigenfunctions:
\begin{equation}
\exp \left [ i {\bf p} \cdot {\bf X} \right ]
 = \sum_{m=0}^{\infty} {{i^m}\over{m!}} ({\bf p}\cdot{\bf X})^m
\quad ,
\label{eq:plwave}
\end{equation}
where each wavefunction is defined by its Taylor series expansion. Such matrix-valued
functions can be manipulated in the usual way as long as we keep in mind the rules for
matrix partial differentiation and matrix integration.

\vskip 0.1in
Let us now construct the tower of extended D-objects that couple to the tower
of higher rank matrix potentials, $C_{[p]}$, in the bosonic matrix theory. We have:
\begin{eqnarray}
X^{\mu_1 \mu_2} =&&  e^{[\mu_1}_{a_1} e^{\mu_2]}_{a_2} d\xi^{a_1} d \xi^{a_2}
\cr
X^{\mu_1 \mu_2 \mu_3 } =&&  e^{[\mu}_{a_1} e^{\mu_2}_{a_2} e^{\mu_3]}_{a_3}
d \xi^{a_1} d \xi^{a_2} d \xi^{a_3}
\cr
\cdots =&& \cdots
\cr
X^{\mu_1 \mu_2 \cdots \mu_{26}} =&& e^{[\mu_1}_{a_1} e^{\mu_2}_{a_2} \cdots
   e^{\mu_{26}]}_{a_26}
d \xi^{a_1} \cdots d \xi^{a_26}
\quad ,
\label{eq:tower}
\end{eqnarray}
coupling, respectively, to matrix potentials $C_{[2]}$, $\cdots$, $C_{[26]}$.
Each is an $N$$\times$$N$ matrix obtained from the tensor product of $p$
adjoint irreps of $SU(N)$. Recall that evidence for the existence of an
ordinary vector potential in some region of space is given by the nonvanishing
holonomy of the gauge potential around a closed loop threaded by the Yang-Mills
field. Likewise, one may verify the existence of a higher rank gauge potential
by performing an integration over a suitable hypersurface in space. It is this
definition which has a nice extension for extended objects within the finite $N$
matrix framework.

\section{The Symmetry Algebra at Finite $N$}

\vskip 0.1in
Having illuminated our admittedly abstract presentation of the bosonic
matrix Lagrangian, it is helpful to return to a clearer discussion of the
symmetries manifest at finite values of $N$. Recall the form of the
Lagrangian from section 2:
\begin{equation}
{\cal L} =
- \fhalf \gap e^{-\Phi} {F}^{\mu\nu} F_{\mu\nu} -
   \half \kap e^{- 2 \Phi} ( {\cal R}
              - 4 \partial^{\mu} \Phi \partial_{\mu} \Phi  ) -
  \thalf \kap  e^{-2 \Phi} {H}^{\mu\nu\lambda} {H}_{\mu\nu\lambda}  \quad .
\label{eq:bmatcc}
\end{equation}
The gauge covariant derivative has already been defined in section 2.3. The Yang-Mills
and antisymmetric threeform field strength may be written more explicitly as follows:
\begin{eqnarray}
F^i_{\mu\nu} =&& \partial_\mu A^i_{\nu} - \partial_\nu A^i_{\mu} +  gf^{ijk} A^j_{\mu} A^k_{\nu}
\cr
H_{\mu\nu\lambda} =&&
\partial_{[\mu} A_{\nu\lambda]} - X_{\mu\nu\lambda} \equiv
\partial_{[\mu} A_{\nu\lambda]} -
  {\rm tr}_{ijk} ~ (\delta_{ij} A^i_{[\mu} F^j_{\nu\lambda]}
  - {{2}\over{3}} f^{ijk} A^i_{[\mu} A^j_{\nu} A^k_{\lambda]} )
\quad .
\label{eq:m3form}
\end{eqnarray}
Notice the definition of the \lq\lq shifted" field strength. With this definition, the
kinetic terms for both $F$ and $H$ take standard form.

\vskip 0.1in
It is helpful to verify explicitly the invariance of the matrix
Lagrangian under a local Lorentz transformation.
Local Lorentz transformations act as tangent space
rotations. We introduce an infinitesimal hermitian matrix, ${\rm L}_{ab}$,
antisymmetric under the interchange of tangent space indices
$a$, $b$. Keeping terms up to linear in ${\rm L}_{ab}$, it is easy
to verify that each of the kinetic terms in ${\cal L}$ is invariant
under the following transformations:
\begin{eqnarray}
\delta e_{a}^{\mu} =&& {\rm L}_a^c e_{c}^{\mu}
\cr
\delta A_a  =&&  {\rm L}_{a}^c  A_c
\cr
\delta F_{ab}  =&&  {\rm L}_{a}^e F_{eb} +  {\rm L}^e_b  F_{ae}
\cr
\delta A_{ab}  =&&  {\rm L}_{a}^e  A_{eb}  +  {\rm L}^e_b A_{ae}
\cr
\delta H_{abc}  =&&  {\rm L}_{a}^e  H_{ebc}
+   {\rm L}^e_b  H_{eac}  +   {\rm L}^e_c  H_{abe}
\label{eq:lorentn}
\end{eqnarray}

\vskip 0.1in
Likewise, we can verify invariance of the matrix Lagrangian under
the Yang-Mills transformation. A Yang-Mills transformation is a
rotation in color space. Locality implies the possibility of
independent rotations for the elements along the diagonal of the
matrix potential $A_{\mu}$. We introduce a $d_G$-plet of infinitesimal
real matrices, $\{ \alpha^j \}$, where $d_G$ is the dimension of the
nonabelian gauge group with hermitian generators $\{ \tau_j \}$.
The $\alpha_j$ are required to take diagonal $N$$\times$$N$ form.
With this restriction, it is easy to verify that each term of the
Lagrangian is invariant under the Yang-Mills transformation
given below:
\begin{eqnarray}
\delta g A_a^j \tau^j  =&& [ D_a , \tau^j \alpha^j ]
\cr
\delta  D_{a} \Phi  =&& i \tau^j \alpha^j D_a \Phi
\cr
\delta F_{ab}  =&& i \tau_j \alpha_j F_{ab}
 \quad .
\label{eq:gaugenab}
\end{eqnarray}
Finally, under a gauge transformation mediated by the two-form potential,
the gauge potentials transform as follows:
\begin{equation}
\delta A_{\mu\nu} = \partial_{[\mu} \zeta_{\nu]} ,
\quad \delta A_{\mu} = - \zeta_{\nu}  \quad .
\label{eq:twoform}
\end{equation}
This is also an invariance of the matrix Lagrangian.

\vskip 0.1in
In the generic curved spacetime background, the symbol \lq\lq $;$" may be
used to denote action of the general covariant derivative including
Christoffel connection, generalizing the arguments given above. The
Riemann curvature scalar may be expressed in the explicit form:
\begin{equation}
{\cal R} [E] = (D_b e^{b\lambda}) (D_a e^a_{\lambda } )
-  (D_a e^{b\lambda}) (D_b e^a_{\lambda})
+  e^{a\lambda} (D_a e^b_{\sigma }) (D_b e^c_{\lambda})   e^{\sigma}_c
-  e^{a\lambda}  (D_a e^c_{\lambda }) e_{c}^{\sigma} (D_b e^b_{\sigma})
\quad .
\label{eq:mcurvy}
\end{equation}
Referring back to the Lagrangian, it is clear that the Einstein term as
written is invariant under both local Lorentz and Yang-Mills transformations
at finite $N$.

\vskip 0.1in
Next, consider expanding about a large $N$ limit of bosonic matrix theory
characterized by spatially-extended objects coupled to a $p$-form potential,
$C_{[p]}$, $p$$\le$$26$, namely, Dbranes.
The matrix $p$-form transformations have the following nontrivial
consequence:
under a tensor $p$-form gauge transformation, all
$p^{\prime}$-form gauge potentials, with $p^{\prime}$$\le$$p$, must
transform non-trivially, including the ordinary Yang-Mills potential.
For example, consider the $3$-form potential, $C_{\mu\nu\lambda}$.
We have:
\begin{equation}
\delta C_{\mu\nu\lambda} = \partial_{[\mu} \zeta_{\nu\lambda]} ,
\quad
\delta C_{\nu\lambda} = \partial_{[\nu} \zeta_{\lambda]} -
                                      \zeta_{\nu\lambda} ,
\quad
\delta C_{\lambda} = - \zeta_{\lambda}
\quad .
\label{eq:extend3}
\end{equation}
The corresponding kinetic term can be written in standard form:
\begin{equation}
{\cal L} = \half {\cal F}_{4} \wedge {\cal F}_{4} ,
\quad
{\cal F}_{\mu\nu\lambda\sigma} = \partial_{[\mu} C_{\nu\lambda\sigma]} -
X_{\mu\nu\lambda\sigma} \quad .
\label{eq:4form}
\end{equation}
where the shifted $4$-form field strength, ${\cal F}_{4}$, is defined as follows:
\begin{equation}
X_{\mu\nu\lambda\sigma} =
  - C_{[\mu\nu} C_{\lambda\sigma]}
  -  A^i_{[\mu} A^i_{\nu} C_{\lambda\sigma]}
  -  f_{ijk} A^i_{[\mu} A^j_{\nu} A^k_{\lambda} C_{\sigma]}
  -  A^i_{[\mu} A^i_{\nu} A^j_{\lambda} A^j_{\sigma]}
\quad .
\label{eq:stre4}
\end{equation}
Is it mandatory that the shift take its most general form inclusive of
coupling to all $p$-form potentials with $p$$\le$$3$? In the case of the
matrix transformations, it is indeed the case: if any one $p$-form charge is
carried by the matrix theory vacuum, it automatically carries all of the
$p$-form charges. The result follows as a consequence of the Lorentz and
Yang-Mills invariance of the quantum theory.

\vskip 0.1in
Since the matrix potentials are noncommuting objects, the $U(N)$ commutator,
$[L_{ab},C_{[p]}]$, is nontrivial for any value of $p$. This implies coupling
to a $(p$$+$$2$)-form potential, and, upon iterating this argument, to the
chain of $(p$$\pm$$2n$)-form potentials. Conversely, the nontrivial $U(N)$
commutator, $[A_1 , C_{[p]}]$, implies coupling to a $(p$$\pm$$1$)-form
potential and, by iteration, to all  $(p$$\pm$$n$)-form potentials. This
observation will be termed matrix Dbrane democracy; it follows from Lorentz
invariance and the presence of gauge fields in the large $N$ continuum limit.
In the special case that the gauge fields are nonabelian, matrix Dbrane
democracy has a beautiful remnant in the form of mixed Chern-Simons terms
in the low energy spacetime action of perturbative string theory. It implies,
in particular, that Dbrane charge conservation must be carefully defined
so as to account for the mixing due to the presence of these terms in the
action \cite{douglas}.

\section{Supersymmetric Matrix Theory}

We begin discussion of our proposal for Matrix Theory by incorporating a
crucial feature absent in previous matrix formulations of String/M theory
\cite{bfss,ikkt,wati}, namely, chirality. We
assign bosonic and fermionic
members of each supersymmetry multiplet to distinct
$U(N)$ representations.
Thus, the gaugino, gravitino, and dilatino belong in
the fundamental $N$-dimensional representation of the $SU(N)$ subgroup, while
their bosonic
superpartners, $A_{\mu}$, $e_{\mu}^a$, belong in the $N$$\times$$N$
adjoint representation.
This is an essential point of
difference from previous conjectures for Matrix Theory: matrix variables in
the fundamental representations of $U(N)$ have appeared in previous
work on matrix formulations of heterotic matrix theory \cite{wati}, but
the fermionic
and bosonic superpartners within any multiplet were chosen to belong in the
same $U(N)$ irrep. In other words, {\em our finite $N$ symmetry algebra is not
simply a supersymmetrization of $U(N)$, and the usual formalism of supergroups
and super-matrix models does not apply}.

\vskip 0.15in
A second point of difference from previous work is that there is no need for a
physical gauge fixing in taking the large $N$ limit. Thus, the number of
supersymmetries at finite $N$ is the same as in the large $N$ limit, namely,
sixteen.
The $SU(N)$ matrix variables carry, in addition,
both Lorentz and nonabelian group indices.
In the discussion that follows, we will denote the
finite-dimensional Yang-Mills group as the generic group $\bf G$, of
rank $r_G$, and dimension $d_G$.
With some guidance from the continuum $N$$=$$1$ supergravity-Yang-Mills
Lagrangian, we can infer the form of the kinetic terms for the given
variables in the matrix Lagrangian. We have:
\begin{eqnarray}
{\cal L} &&=~
- \half \kap e^{-2\Phi}
\left ( {\bar \psi}_{\mu} \Gamma^{\mu\nu\lambda} {\cal D}_{\nu} \psi_{\lambda}
~-~ 4 {\bar{\lambda}} \Gamma^{\mu\nu} {\cal D}_{\mu} \psi_{\nu}
~-~ 4 {\bar{\lambda}} \Gamma^{\mu} {\cal D}_{\mu} \lambda \right )
~-~ \half \gap e^{-\Phi}~ {\bar \chi}^i \Gamma^{\mu} {\cal D}_{\mu} \chi^i
\cr
\quad\quad  &&\quad\quad
~-~ \half \kap e^{-2\Phi} \left ( {\cal R}
~-~ 4~ \partial^{\nu} \Phi ~ \partial_{\nu} \Phi \right )
~-~  \thalf \kap e^{-2 \Phi} ~ {H}^{\mu\nu\lambda} {H}_{\mu\nu\lambda}
~-~ \quart \gap e^{-\Phi }~ {F}^{\mu\nu} F_{\mu\nu}
\cr
\quad && \quad\quad\quad\quad
~+~   {\cal L}_{\rm 2-fermi}  ~+~ {\cal L}_{\rm 4-fermi}  \quad ,
\label{eq:hmat1}
\end{eqnarray}
where the two- and four-fermi terms will be inferred by requiring
closure under the supersymmetry transformations.
In the expression above, $\chi^{i\alpha} $, ${\bar{\chi}}^{i\alpha}$,
denote Grassmann-valued fermionic matrices in the $N$, ${\bar{N}}$,
representations of $SU(N)$. The indices, $i$$=$$1$, $\cdots$, $r_G$,
simultaneously labels a fundamental representation of the Yang-Mills group
$\bf G$, while $\alpha$$=$$1$, $\cdots$, $16$, labels sixteen distinct
Grassmann-valued fermionic matrices. The spinor covariant derivative is
both Lorentz, and gauge, covariantized:
\begin{equation}
{\cal D}_{\mu} \chi = \partial_{\mu} \chi
 + \half \Gamma^{ab} \Omega_{\mu ab} \chi + g A_{\mu} \chi \quad ,
\label{eq:cov}
\end{equation}
where $\Omega_{\mu ab}$ is the three-index spin connection \cite{fs,br}.
In the large $N$ limit, these are diagonal matrices, corresponding to
the sixteen components of a Majorana-Weyl
spinor field. In the proper time gauge, distance along the diagonal
has been mapped to time, and each diagonal element is a smooth function
of the spatial coordinates. Thus, we recover the compoents of a
Grassmann field.

\vskip 0.1in
Likewise, $\psi^{\alpha}_{\mu }$, denotes Grassmann-valued fermionic
matrices evolving in the large $N$ limit into the components of a
Lorentz spinor-vector field in ten dimensions.
Finally, we have the matrix representatives of the dilatino field,
also living in a Grassmann-valued $SU(N)$ fundamental representations,
$\lambda^{\alpha}$. In the continuum limit, $\chi^i$,
$\psi_{\mu}$, and $\lambda$, yield, respectively, the
gaugino, gravitino, and dilatino fields of the d=10
${\cal N}$$=$$1$ SYM supergravity Lagrangian. The $SU(N)$ matrices
$F_{ab}$, $H_{abc}$, $\cal R$, and $\Phi$ are, respectively,
finite $N$ matrix representatives of
the Yang-Mills tensor, the shifted antisymmetric
three-form field strength corresponding to the
two-form potential $C_{[2]}$, plus Chern-Simons term for the
Yang-Mills potential, the Ricci
curvature, and the dilaton scalar continuum fields.

\vskip 0.1in
Closure of the group of transformations that are the finite
$N$ manifestation of large $N$ continuum supersymmetry algebra
is a nontrivial result. However, as we will see below, with the
ordering prescription given earlier, the manipulations required
to verify that $\cal S$ is supersymmetry invariant are well-defined.
Consider infinitesimal spinor parameters, $\eta_1$, $\eta_2$, each
of which transforms as a $N$-vector of the
unitary group $SU(N)$. We must verify that the commutator
of two matrix supersymmetry transformations
with arbitrary infinitesimal spinor parameters
can always be expressed as the sum of
(i) an infinitesimal tangent space translation with parameter,
$\xi^{a}$$=$${\bar{\eta_1}} \Gamma^{a}
\eta_2$, (ii) an infinitesimal
local Lorentz transformation with parameter
${\rm L}_{bc}$$=$$\xi^{a} \omega_{abc}$, and (iii)
an infinitesimal local gauge transformation with gauge
parameter $\alpha^i$$=$$-g \xi^{a} A_{a}^i$ \cite{fs}.

\vskip 0.15in The form of the supersymmetry transformations is as
follows. We consider the following sequence of matrix
transformations induced by the infinitesimal spinor parameter,
$\eta$, a Grassmann-valued, $N$-dimensional vector under $SU(N)$:
\begin{eqnarray}
\delta e^a_{\mu} &&= \half {\bar{\eta}} \Gamma^a \psi_{\mu}
\cr
\delta \Phi &&=  - \half {\bar{\eta}} \lambda
\cr
\delta A_{\mu\nu} &&= \half {\bar{\eta}} \Gamma_{[\mu} \psi_{\nu]}
- \gap {\rm tr} (A_{[\mu} \delta A_{\nu]} )
\cr
\delta A^i_{\mu}  &&= \half {\bar{\eta}} \Gamma_{\mu} \chi^i
\cr
\delta \psi_{\mu} &&= {\cal D}_{\mu} \eta + \half \left (
{\bar{\eta}} \psi_{\mu} - {\bar{\psi}}_{\mu} \eta \right ) \lambda
 - \half  ( {\bar{\psi}}_{\mu} \Gamma^a \eta ) \Gamma_a \lambda
+ \ninsix \gap {\rm tr} ({\bar{\chi}} \Gamma_{abc} \chi ) \Gamma^{abc} \Gamma_{\mu} \eta
\cr
\delta \lambda &&=  -  \half (\Gamma^a {\cal D}_a \Phi  ) \eta +
\quart \left ( {\hat{H}}_{abc} - \twff {\bar{\lambda}} \Gamma_{abc} \lambda
+ \ninsix \gap  {\rm tr} ( {\bar{\chi}} \Gamma_{abc} \chi ) \right )
 \Gamma^{abc} \eta
\cr
\delta \chi^i &&= - \quart  ( \Gamma^{ab} {\hat{F}}^i_{ab} ) \eta
+ \half \left ( {\bar{\eta}} \chi^i - {\bar{\chi}}^i \eta \right ) \lambda
- \half ({\bar{\chi}}^i \Gamma^a \eta ) \Gamma_a \lambda
\quad .
\label{eq:susyt}
\end{eqnarray}
It may be verified that there is no ambiguity in the ordering of variables
in the transformation laws given here.
We then complete our expression for the Matrix Theory Lagrangian by including
the two-fermion and four-fermion terms required by supersymmetry \cite{fs,br}. With
guidance from the continuum $N$$=$$1$ supergravity-Yang-Mills
Lagrangian \cite{br}, we infer the following 2-fermi terms:
\begin{eqnarray}
 {\cal L}_{\rm 2-fermi} &&=
 - \kap e e^{-2\Phi} {\bar{\psi}}_{\mu} \Gamma^{\mu} \psi_{\nu} ( \partial^{\nu} \Phi )
~+~ 2 \kap e e^{-2\Phi} ~{\bar {\psi}}_{\mu} \Gamma^{\nu} \Gamma^{\mu} \lambda ( \partial_{\nu} \Phi)
\cr
&&\quad\quad + \eigt \kap e e^{-2\Phi} H^{\rho\sigma\tau} \left [
{\bar \psi}_{\mu} \Gamma^{[\mu} \Gamma_{\rho\sigma\tau }\Gamma^{\nu]} \psi_{\nu}
~+~ 4 ~{\bar \psi}_{\mu} \Gamma^{\mu}_{\rho\sigma\tau } \lambda
~-~ 4 ~ {\bar\lambda} \Gamma_{\rho\sigma\tau } \lambda \right ]
\cr
&&\quad\quad\quad
~+~ \eigt \gap e e^{-\Phi} {\hat{H}}^{abc} {\rm tr} ({\bar \chi} \Gamma_{abc} \chi )
~-~ \eigt \gap e e^{-\Phi} ~ {\bar \chi}^i \Gamma^{\mu} \Gamma^{ab}
(\psi_{\mu} +  \thre \Gamma_{\mu} \lambda ) ( F_{ab}^i + {\hat{F}}_{ab}^i )
\quad .
\label{eq:2fermat}
\end{eqnarray}
The 4-fermi terms in the Lagrangian take the form:
\begin{eqnarray}
{\cal L}_{\rm 4-fermi} &&=
- \nineig \kap e e^{-2\Phi} {\bar\psi}^{\mu} \Gamma^{abc} \psi_{\mu}
 \left (  {\bar\psi}_{\nu} \Gamma^{\nu} \Gamma_{abc} \Gamma^{\lambda} \psi_{\lambda}
      + 2 ~ {\bar{\psi}}^{\nu} \Gamma_{abc} \psi_{\nu}
        - 4 ~ {\bar{\lambda}} \Gamma_{abc} \lambda
          - 4~ {\bar{\lambda}} \Gamma_{abc} \Gamma^{\nu}\psi_{\nu}
                   \right )
\cr
&&\quad\quad
~+~ \ninsix \gap e e^{-\Phi} ~ {\rm tr} ( {\bar{\chi}} \Gamma^{abc} \chi ) \left (
- \half {\bar{\psi}}_{\mu} ( 4~ \Gamma_{abc} \Gamma^{\mu} + 3 ~ \Gamma^{\mu} \Gamma_{abc} ) \lambda
   + {\bar{\lambda}}\Gamma_{abc}\lambda -  24 ~ {\hat{H}}_{abc} \right )
\cr
&&\quad\quad\quad\quad
- \ninsixp \gap e e^{- 2\Phi}
{\rm tr} ( {\bar{\chi}} \Gamma^{abc} \chi)\cdot \gap {\rm tr}( {\bar{\chi}} \Gamma_{abc} \chi )
 \quad .
\label{eq:4fermimat}
\end{eqnarray}
The expression for ${\cal L}$ may be simplified and written even more compactly
by introducing $SU(N)$ vectors, $\Psi$, ${\bar{\Psi}} $,
$(d$$+$$1$$+$${\rm d}_{\rm G})$$\times $$N$-component $SU(N)$
vectors. Each transforms simultaneously as, respectively, 16-component
right- and left-handed
Majorana-Weyl spinors under the inhomogenous Lorentz group. They are
denoted as follows:
\begin{equation}
{\bar{\Psi}} \equiv ({\bar{\lambda}} , {\bar{\psi}}_{a} , {\bar{\chi}}^i )  ,
\quad \quad
\Psi \equiv (\lambda , \psi_{b} , \chi^j ) \quad .
\label{eq:fermions}
\end{equation}
The independent Lorentz structures present in the kinetic and
two-fermi terms of ${\cal L}$ may be grouped inside a matrix array of size
$(d$$+$$1$$+$${\rm d}_{\rm G})N $$\times $$ (d$$+1$$+$${\rm d}_{\rm G})N $,
which we denote as ${\cal D}$.
The four-fermi terms are likewise expressed in compact form
by introducing matrices, $\cal U$, $\cal V$, of size
$(d$$+$$1+{\rm d_G})N$$\times$$(d$$+$$1+{\rm d_G})N$, identified by referring
to the expression in Eq.\ (\ref{eq:4fermimat}). In summary, the classical
Lagrangian for Matrix Theory takes the remarkably compact form:
\begin{equation}
{\cal L} ~=~ - \half {\bar\Psi} {\cal D} \Psi +
  \quart ({\bar{\Psi}}{\cal U} \Psi )( {\bar{\Psi}} {\cal V} \Psi)
  - \quart \gap  e^{-\Phi} ~ F^{\mu\nu} F_{\mu\nu}
  - \half \kap e^{-2\Phi} ~ ( {\cal R}
     - 4 \partial^{\mu} \Phi \partial_{\mu}  \Phi
     +3  H^{\mu\nu\lambda} H_{\mu\nu\lambda} )
\quad .
\label{eq:compact}
\end{equation}
We should note that, in principle, ${\cal L}$ belongs to a family of
matrix Lagrangians, members of which can differ by $1/N$ corrections, thus yielding
the same spacetime Lagrangian in the infrared in accordance with the principle of
universality classes. However, we can state definitively that the universality class
of our theory does not overlap with either the BFSS or IKKT matrix models because
of the distinct $U(N)$ assignments given to the members of a supermultiplet in our
framework. Our procedure for determining $\cal L$ ensures that all
relevant interactions in the large $N$ continuum Lagrangian that are required in order
to match correctly with a spacetime Lagrangian that is manifest Yang-Mills invariant,
locally supersymmetric, and Lorentz invariant at the scale $\alpha^{\prime -1/2}$,
are already present in the ultraviolet theory defined by ${\cal L}$. Thus, the sole
source for both nonperturbative and quantum corrections to the spacetime Lagrangian
are the quantum corrections from the matrix path integral.

\section{The Wheeler De Witt Equation}

It should be possible to derive a Wheeler-De Witt equation for
Matrix Theory as follows. We will specialize to proper time gauge, where we
identify $X^{0}$$\equiv$$E^{0}_a (X) d \xi^a$ with tangent space time,
$X^0$$=$$\xi^0$,
at all points in space, which remains discrete. This implies setting the
$E^0_a$ to zero for all spatial $a$, and $E^0_0$$=$$1$, the unit $N$$\times$$N$
matrix. We will work in Euclidean time, and the end-points, $n$$=$$1$, $N$,
correspond to the box-regularization, $X^0$$=$$0$, $T$, We will formulate the
Wheeler-DeWitt equation for the matrix quantum mechanics thus defined, enabling
construction of the Hartle-Hawking wavefunction \cite{hh,sh}. For convenience,
we rename the timelike coordinate, $X^0$$=$$\xi^0$$=$$t$.

\vskip 0.1in
Matrix quantum dynamics in the proper time gauge is given by the Schroedinger equation
\cite{hh}, $i \partial \Psi /\partial t$ $=$ $H\Psi$.
The wavefunction for the ground state, or state of minimum excitation, $\Psi_0$,
is defined by the matrix
path integral, made positive definite by a rotation to Euclidean time.
As explained in
\cite{hh,sh}, even though there is strictly speaking no minimum energy state in
a theory of quantum gravity, our gauge fixing condition makes both the notion
of energy and of the minimum energy state well-defined. At the initial
time $t$$=$$0$, we have:
\begin{equation}
\Psi_0 [e^{m}_a (0) ; \phi(0) ] = \int d [ e^m_a (t)]d[ \phi ]
    \exp \left \{ -I [e_a^m ; \phi ] \right \}
\quad ,
\label{eq:grd}
\end{equation}
where $I$ is the Euclidean action given by the variation of the matrix
action described in the previous section.
The $e^m_a (0) $ specify a particular spatial nine-geometry, and the wavefunction
is the amplitude for that geometry to be created from the zero nine-geometry---
a single point, or nothing, at the initial time \cite{hh}. The matrix matter and
gauge degrees of freedom are the additional data that must be specified on the initial
value slice in discrete spacetime, namely, on the spatial nine-geometry valid at
$t$$=$$0$. The matter and gauge degrees of freedom, $A$, $C_{[p]}$, $\phi$,
$\chi$, $\psi$, $\lambda$, have been collectively denoted by the symbol $\phi$.

\vskip 0.1in
In practice, an object of more direct interest is the probability of finding
a certain closed, compact submanifold, $S$, with given 9-geometry and given
configuration of regular matter fields, and which divides the spacetime manifold
into in- and out- manifolds, $M_{\pm}$. Such a probability can be factorized into
the product of amplitudes, $\Psi_{\pm}$, where the path integral sums over classes
$C_{\pm}$, of 9-geometries and matter fields on $M_{\pm}$, which match with the
given 9-geometry on $S$. Following \cite{hh,sh}, the $\Psi_{\pm}$ may be regarded
as wave functions of the Universe. If the classes, $C_{\pm}$, are identical,
we can drop the suffix $\pm$ without ambiguity. The functional differential
equation satisfied by $\Psi$ is the Wheeler-De Witt equation. In the case of
Matrix Theory, this is defined by introducing a proper time which is constant on
$S$, giving the standard lapse-shift decomposition of the $9$$+$$1$-metric. The
Wheeler De Witt equation is the matrix functional differential equation obtained
by varying the classical matrix action with respect to the lapse function. We
will not obtain its precise form in this paper, but we make the following remarks.

\vskip 0.1in
In \cite{hh,sh}, a major focus of interest is the issue of what constitutes the
class of spatial geometries that must be summed over in the path integral. The
argument is made that it is the no-boundary geometries based on compact, positive
definite metrics that are relevant to the wavefunction of the Universe and hence
to quantum cosmology.
It is interesting to contrast this with the case of two-dimensional
gravity \cite{polg}, namely, first quantized string theory, where the path integral localizes
on a finite dimensional integral and there is no ambiguity about the class of
geometries summed in the path integral. The reason for this is Weyl invariance;
any metric in two dimensions is gauge equivalent to a constant curvature metric,
leaving only a finite dimensional integral over the worldsheet moduli.

\vskip 0.1in
Although we will not aim to settle this thorny issue in this paper, we would like
to propose that the same is true in a fundamental theory of the Universe. Namely,
Matrix Theory is both a dynamical theory {\em and} a theory of the ground state \cite{hart1}.
The configuration space of spacetime geometries and background fields leading to a finite, and
renormalizable, perturbation theory in the infrared should define a complete set, selected
by the high degree of symmetry of the solutions, both kinematic and dynamic. Any other
background geometry will be gauge equivalent to a member of this set, where by a gauge
equivalence here we mean a transformation falling under the category of spacetime
$\times$ internal symmetry transformations. Thus, the higher rank quantum gauge
invariances become essential to the understanding of the full configuration space of
the matrix path integral. In the large $N$ continuum limit, the quantum gauge symmetries
manifest themselves as the strong-weak and target space dualities linking the different
low energy limits of M theory \cite{mthy,dvv,kth,malda}.

\vskip 0.1in
We should emphasize, however, that we are in broad agreement with the arguments
put forth in \cite{hh,sh} that asymptotically flat and AdS geometries, while
natural in particle physics with its focus on a framework suited to scattering,
and also widespread in perturbative string theory \cite{malda}, are too limiting
a class of geometries of likely relevance to cosmology. Fortunately, our
understanding of de Sitter-like spacetimes in String/M theory is rapidly undergoing
development \cite{desit}
and there is hope that one will have a clearer perspective on this subject
in the future. We reiterate that it is our hope that the choice of boundary condition
in Matrix Theory will have an unambiguous origin as in perturbative string theory,
without recourse to an independent principle originating outside of the theory.

\section{Conclusions}

\vskip 0.1in
Our proposal for Matrix Theory is reminiscent of a discretization of the
$\alpha^{\prime}$ expansion of the spacetime Lagrangian for String Theory.
The $\alpha^{\prime}$ expansion is nonperturbative in the coupling constant
and, not surprisingly, has been a major source of insight into strongly
coupled string theory and its nonperturbative solutions. However, the
discretization or, more precisely, regularization offered by the matrix
description is not to be confused with ordinary lattice field theories. The
number of degrees of freedom associated with each of the coordinates of space
is $N^2$, rather than the expected $N$, and the spacetime geometry is
noncommutative. Points in spacetime are in one-to-one correspondence with
matrices, and by introducing the auxiliary device of a continuum tangent space,
we have achieved a diffeomorphism invariant description capable of accomodating
arbitrary curved spacetime geometries. These are significant gains although
the resulting matrix Lagrangian is understandably complex. It should be
emphasized, however, that the bosonic Lagrangian is relatively simple, and
the fermionic additions to it are mostly a matter of achieving closure of the
supersymmetry algebra.

\vskip 0.1in
A second theme running through this work has been the notion of gauge
symmetry, both classical and quantum. We have stressed the role of the
higher rank antisymmetric Lorentz tensors which couple to extended objects,
pointing out how, at the level of the finite $N$ algebra, charge under any one
gauge potential implies charge under the full tower of potentials.
In the large $N$ continuum limit, the quantum gauge symmetries manifest
themselves as strong-weak and target space dualities linking the different
low energy limits of M theory. The notion of Duality as a gauge symmetry is
not a new idea, and has already received considerable attention in the
literature. Our work can be taken as further evidence for the validity of
this notion. The principle of quantum gauge invariance will be of fundamental
importance in any precise treatment of the matrix path integral.

\vskip 0.1in
However, we expect that, as with the $\alpha^{\prime}$ expansion of string
theory, the immediate most fruitful directions of work will come from semi-classical
analyses of Matrix Theory. The derivation of the Wheeler--De Witt equation, and
the possibility of studying quantum cosmology that it opens up, are of the greatest
interest here. We leave that effort for future work.

\newpage
\centerline{\bf ACKNOWLEDGMENTS}

\vskip 0.1in
\noindent
This work was completed in part during visits at Kyoto University, Cornell
University, Institute for Theoretical Physics (Santa Barbara), Ecole Normale
Superior (Paris), Service de Physique Theorique (Saclay), University of
Cambridge, and the Amsterdam Summer Workshop. Their hospitality is gratefully
acknowledged.

\vskip 0.4in 
\noindent{\bf Note Added (July 2005):} This paper continues the stream of
conceptual advances in the development of my proposal for nonperturbative 
String/M theory from hep-th/0201129, 0202138, and 0205306. The notion of 
emergent spacetime
as introduced by me first appears in this paper. For a clearer presentation, the
reader should consult hep-th/0408057 [Nucl. Phys B719 (2005) 188].

\end{document}